\documentclass[a4paper,12pt]{article}
\usepackage{latexsym}
\usepackage{graphicx}
\usepackage{dcolumn}
\usepackage{bm}
\usepackage[T1]{fontenc}
\usepackage[ansinew]{inputenc}
\usepackage{amsmath}
\usepackage{amssymb}
\usepackage{graphicx}
\begin{document}

\title{PARADOX OR NON-PARADOX IN WAVE-PARTICLE DUALITY }
\date{\today}
 \maketitle
\textbf{A.~Drezet}\\
\emph{Institut N\'{e}el, CNRS-Universit\'{e} Joseph Fourier, UPR 2940, 25 rue des Martyrs, 38000, Grenoble, France }\\
\emph{email: aurelien.drezet@grenoble.cnrs.fr}\\
\date{Received :}\\
We analyze the experiment recently realized by S.~Afshar
\emph{et al.} \cite{Afshar} in order to refute the principle of complementarity. We discuss the general meaning of this principle and show that
contrarily to the claim of the authors Bohr's complementarity is not in danger in this
experiment.
\newpage
\section{Introduction}
In a recent and interesting article S.~Afshar and
coworkers~\cite{Afshar} (see also \cite{Afshar2}) reported an
optical experiment in which they claimed to refute the well known
N.~Bohr principle of
complementarity~\cite{Bohr1,Bohr2,Bohr2b,Bohr3}. Obviously this
result, if justified, would constitute a serious attack against
the orthodox interpretation of quantum mechanics (known as the
Copenhagen interpretation). This work stirred much debate in
different journals (see for examples references
\cite{A,B,C,D,E,F,G,H}). We think however that there are still
some important misunderstandings concerning the interpretation of
this experiment. In a preprint written originally in 2004 \cite{A}
(and following some early discussions with Afshar) we claimed
already that the interpretation by Afshar \emph{et al.} can be
easily stated if we stay as close as possible from the texts
written by Bohr. The aim of the present article (which was
initially written in 2005 to precise a bit the thought developed
in \cite{A}) is to comment the interpretation discussed in
\cite{Afshar}. We will in the following analyze the meaning of
Bohr principle and show that far from disproving its content the
experiment~\cite{Afshar} is actually a complete confirmation of
its
general validity.\\
\indent  The difficulties associated with the understanding of
this principle are not new and actually complementarity created
troubles even in Einstein mind~\cite{Bohr1} so that we are here in
good company. To summarize a bit emphatically Bohr's
complementarity we here remind that this principle states that if
one of a pair of non commuting observables of a quantum object is
known for sure, then information about the second (complementary)
is lost \cite{Bohr1,Bohr2,
Bohr3,Heisenberg1,Heisenberg2,Heisenberg3,Feynman,Zeilinger}. This
can be equivalently expressed as a kind of duality between
different descriptions of the quantum system associated with
different experimental arrangements which mutually exclude each
other (read in particular \cite{Bohr1,Bohr2,Bohr3}). Later in the
discussion we
will try to precise this definition but for the moment it is enough to illustrate the concepts by examples\\
\indent Consider for instance the well known Young double-pinholes
interference experiment made with photons. The discrete nature of
light precludes the simultaneous observation of a same photon in
the aperture plane and in the interference pattern: the photon
cannot be absorbed twice. This is already a trivial manifestation
of the principle of Bohr. Here it implies that the two statistical
patterns associated with the wave in the aperture plane and its
Fourier (i.~e., momentum) transform require necessarily different
photons for their recording. It is in that sense that each
experiment excludes and completes reciprocally the other. In the
case considered before the photon is absorbed during the first
detection (this clearly precludes any other detection). However
even a non-destructive solution for detection implying
entanglement with other quantum systems has a radical effect of
the same nature: the complementarity principle is still valid. For
example, during their debate Bohr and Einstein \cite{Bohr1}
discussed an ideal \emph{which-way} experiment in which the recoil
of the slits is correlated to the motion of the photon. Momentum
conservation added to arguments based on the uncertainty relations
are sufficient to explain how such entanglement photon-slits can
erase fringes~\cite{Feynman,Zeilinger,Scully,Drezet1,Drezet2}.\\
\indent It is also important for the present discussion to remind that the principle of
complementarity has a perfidious consequence on the experimental
meaning of trajectory and path followed by a particle. Indeed the
unavoidable interactions existing between photons and detectors
imply that a trajectory existing independently of any measurement
process cannot be unambiguously defined. This sounds even like a
tragedy  when we consider once again the two-holes experiment.
Indeed for Bohr this kind of experiments shows definitely the
essential element of ambiguity which is involved in ascribing
conventional physical attributes to quantum systems. Intuitively
(i.~e., from the point of view of classical particle dynamic) one
would expect that a photon detected in the focal plane of the lens
must have crossed only one of the hole 1 or 2 before to reach its
final destination. However, if this is true, one can not intuitively
understand how the presence of the second hole (through which the
photon evidently did not go) forces the photon to participate to an
interference pattern (which obviously needs an influence coming from
both holes). Explanations to solve this paradox have been proposed
by de Broglie, Bohm, and others using concepts such as empty waves
or quantum potentials \cite{Broglie,Bohm}. However all these
explanations are in agreement with Bohr principle (since they fully reproduce quantum predictions) and can not be experimentally
distinguished. Bohr and Heisenberg proposed for all needed purposes
a much more pragmatic and simpler answer: \emph{don't bother}, the
complementarity principle precludes the simultaneous observation of
a photon trajectory and of an interference pattern. For Bohr
\cite{Bohr1}: \emph{This point is of great logical consequence,
since it is only the circumstance that we are presented with a
choice of either tracing the path of a particle or observing
interference effects, which allows us to escape from the paradoxical
necessity of concluding that the behaviour of an electron or a
photon should depend on the presence of a slit in the diaphragm
through which it could be proved not to pass.} From such an analysis
it seems definitively that Nature resists to deeper experimental
investigation of its ontological level. As summarized elegantly by
Brian Greene \cite{Greene}: \emph{Like a Spalding Gray soliloquy, an
experimenter's bare-bones measurement are the whole show. There
isn't anything else. According to Bohr, there is no backstage}. In
spite of its interest it is however not the aim of the present
article to debate on the full implications of such strong
philosophical position.

\section{Complementarity versus the experiments}
\subsection{ A short description of the Afshar \emph{et al.} experiment}
The experiment reported in \cite{Afshar} (see Fig.~1) is actually
based on a modification of a \emph{gedanken} experiment proposed
originally by Wheeler~\cite{Wheeler}. In the first part of their
work, Afshar \emph{et al.} used an optical  lens to image the two
pinholes considered in the Young interference experiment above
mentioned. Depending of the observation plane in this microscope
we can then obtain different complementary information.
\begin{figure}[h]
\begin{center}
\includegraphics[width=10cm]{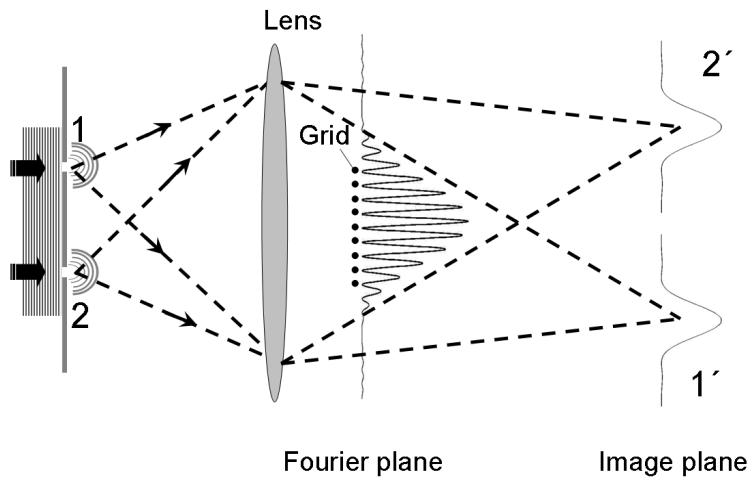}
\end{center}
\caption{The experiment described in \cite{Afshar}. Photons coming
from pinholes 1 and 2 interfere in the back-focal plane of a lens
(Fourier plane) whereas they lead to two isolated narrow spots in
the image plane (the image plane is such that its distance $p'$ to
the lens is related to the distance $p$ between the lens and the
apertures screen by $1/p+1/p'=1/f$, where $f$ is the focal
length). The wire grid in the back focal plane, distant of $f$
from the lens, is passing through the minima of the interference
pattern. The subsequent propagation of the wave is consequently
not disturbed by the grid. }
\end{figure}
If we detect the photons in the focal plane of the lens (or
equivalently just in front of the lens \cite{Remark1}) we will
observe, i.e, after a statistical accumulation of photon detection
events, the interference fringes. However, if we record the
particles in the image plane of the lens we will observe (with a
sufficiently high numerical aperture) two sharp spots 1' and 2'
images of the pinholes 1 and 2. Like the initial Young two-holes
experiment this example illustrates again very well the principle
of Bohr. One has indeed complete freedom for measuring the photon
distribution in the image plane instead of detecting the fringes
in the back focal plane. However, the two kinds of measurements
are mutually exclusive: a single photon can
participate only to one of these statistical patterns.\\
\indent In the second  and final part of the experiment, Afshar
\emph{et al}., included a grid of thin absorbing wires located in
the interference fringes plane. Importantly, in the experiment the
wires must be located at the minimum of the interference pattern
in order to reduce the interaction with light. In the following we
will consider a perfect interference profile (with ideal unit
visibility
$V=(I_{\texttt{max}}-I_{\texttt{min}})/(I_{\texttt{max}}+I_{\texttt{min}})=1$)
to simplify the discussion. If additionally the geometrical cross
section of each wire tends ideally to zero then the interference
behavior will, at the limit, not be disturbed and the subsequent
wave propagation will be kept unchanged. This implies that the
photon distributions 1' and 2', located in the image plane
optically conjugated with the aperture plane, are not modified by
the presence, or the absence,
of the infinitely thin wire grid.\\
\indent Naturally, from practical considerations an infinitely thin
dielectric wire is not interacting with light and consequently
produces the same (null) effect whatever its location in the light
path (minimum or maximum of the interference for example). In
order to provide a sensible probe for the interference pattern,
necessary for the aim of the experiment considered, we will
suppose in the following idealized wires which conserve a finite
absorbtion efficiency and this despite the absence of any
geometrical transversal extension. We will briefly discuss later
what happens with spatially extended scattering wires with finite
cross section, but this point is not essential to understand the
essential of the argumentation. With such wires, and if we close
one aperture (which implies that there is no interference fringes
and thus that a finite field impinges on the wires) the scattering
and absorbtion strongly affect the detection behavior in the image
plane. As it is seen experimentally \cite{Afshar,Afshar2} the
scattering by the wire grid in general produces a complicated
diffraction pattern and not only an isolated narrow peak in 1' or
2' as it would be without the
grid . \\
\indent In such conditions, the absence of absorbtion by the wires
when the two apertures are open is a clear indication of the
existence of the interference fringes zeros, i.e., of a wave-like
character, and this even if the photon is absorbed in the image
plane in 1' or 2'. Following Afshar \emph{et al.}, this should be
considered as a violation of complementarity since the same
photons have been used for recording \emph{both} the `path' and
the wave-like information. The essential questions are however
what we mean precisely here by path and wave-like information and
what are the connections of this with the definition of
complementarity. As we will see hereafter it is by finding a clear
answer to these questions that the paradox and the contradictions
with Bohr's complementarity are going to vanish.
\subsection{The wave-particle duality mathematical relation}
\indent At that stage, it is important to point out that the
principle of complementarity is actually a direct consequence of
the mathematical formalism of quantum mechanics and of its
statistical interpretation (see for example W.~Heisenberg
\cite{Heisenberg3}). It is in particular the reason why the
different attempts done by Einstein to refute complementarity and
the Heisenberg uncertainty relations always failed: the
misinterpretations resulted indeed from a non-cautious
introduction of classical physics in the fully consistent quantum
mechanic formalism. Similarly here we show that a problem of the
same nature occurs since Afshar \emph{et al.} actually mixed
together, and imprudently, argumentations coming from classical
and quantum physics. We will show that this mixing results into an
apparent refutation of the
complementarity principle.\\
\indent After this remark we now remind that an, apparently,
simple mathematical formulation of complementarity exists in the
context of two path interferometry
\cite{Englert1,Englert2,Englert3,Englert4}. For example in the
Young double-apertures experiment considered previously the field
amplitudes $C_1$ and $C_2$ associated with the two narrow
apertures, separated by the distance $d$, allow us to define the
wave function in the two-apertures plane by:
\begin{eqnarray}
\psi(x)\sim C_1\delta(x-d/2)+C_2\delta(x+d/2).
\end{eqnarray}
From this formula one can easily introduce the
`distinguishability'
\begin{eqnarray}
K=\frac{||C_1|^2-|C_2|^2|}{|C_1|^2+|C_2|^2}.
\end{eqnarray}
This quantity can be physically defined by recording the photons
distribution in the aperture plane and constitutes an observable
measure of the path distinguishability \cite{Remark2}. The
interpretation of $K$ is actually clear, and in particular if
$K=0$ each apertures play a symmetrical role, whereas if $K=1$ one
of the two apertures is necessarily closed. Naturally, like in the
Afshar experiment, $K$ can also be measured by recording photons
in the image plane of the lens in 1' and 2'. Equations (1) and (2)
are still valid, with the only differences that:  i) we have now a
diffraction spot (like an Airy disk) instead of a Dirac
distribution in equation (1), and  ii) that the spatial variables
are now
magnified by the lens~\cite{Born}.\\
\indent Instead of the spatial representation one can also
consider the Fourier transform corresponding to the far field
interference pattern recorded at large distance of the two-slits
screen:
\begin{eqnarray}
\psi(k)\sim C_1 \cdot e^{ikd/2}+C_2 \cdot e^{-ikd/2}.
\end{eqnarray}
Such a wave is associated with an oscillating intensity in the
k-space given by \begin{equation}I(k)\sim
1+V\cos{(kd+\chi)}\end{equation} where
$\chi=\arg{(C_1)}-\arg{(C_2)}$ and $V$ is the fringe visibility
\begin{equation}
V=\frac{2|C_1|\cdot|C_2|}{|C_1|^2+|C_2|^2}.
\end{equation}
This quantity is also a physical observable which can defined by
recording the photons in the far-field, or, like in the Afshar
\emph{et al.} first experiment, by recording the photons fringes
in the back focal plane of the lens (the back focal plane is the
plane where the momentum distribution $\hbar k$ is experimentally
and rigorously defined~\cite{Born,Zeilinger}). Like it is for $K$,
the meaning of $V$ is also very clear: if $V=1$ both apertures
must play a symmetrical role, whereas if $V=0$ only one aperture
is
open.\\
\indent  A direct mathematical consequence of equations (2) and
(5) is the relation
\begin{eqnarray}
V^2+K^2=1,
\end{eqnarray}
which expresses the duality \cite{Englert1,Englert2} between the
two mathematical measures $K$ and $V$ associated with the two
mutually exclusive (i.e., complementary) experiments in the direct
and Fourier space respectively. A particularly important
application of equation (6) concerns which-path experiments. In
such experiments, we wish to observe the interference pattern, and
to find through each hole each photon is going through. As we
explained before, a photon can not be observed twice, and this
represents in general a fatal end for such expectations. There is
however an important  exception in the particular case with only
one aperture open (i.e., $K=1$). Indeed, in such case it is not
necessary to record the photon in the aperture plane to know its
path since if it is detected (in the back focal plane) it
necessarily means that it went through the opened aperture. Of
course, from equation (6) we have in
counterpart $V=0$, which means that fringes are not possible.\\
\indent This dilemma, can not be solved by considering less
invasive methods, like those using entanglement between the photon
and an other quantum system or an internal degree of freedom (such
as polarization or spins). To see that we consider a wave function
$|\Psi\rangle$ describing the entanglement between the photon and
these others quantum variables defining a which-path detector. We
write
\begin{eqnarray}
|\Psi\rangle=\int[C_1\delta(x-d/2)|x\rangle|\gamma_1\rangle+C_2\delta(x+d/2)|x\rangle|\gamma_2\rangle]dx\nonumber\\
=\int[C_1\cdot e^{ikd/2}|k\rangle|\gamma_1\rangle+C_2\cdot
e^{-ikd/2}|k\rangle|\gamma_2\rangle]dp
\end{eqnarray}
where $|\gamma_1\rangle$ and $|\gamma_2\rangle$ are the quantum
state of the which path detector if the photon is going through
the aperture $1$ or $2$. Consider now the kind of information one
can extract from $|\Psi\rangle$. First, by averaging (tracing)
over the detector degrees of freedom we can define the total
probability of detecting a photon in the aperture plane in $x$ by
\begin{eqnarray}
P(x)=\textrm{Tr}[\hat{\rho}|x \rangle\langle x|]\propto
|C_1|^2\langle\gamma_1|\gamma_1\rangle(\delta(x-d/2))^2+|C_2|^2\langle\gamma_2|\gamma_2\rangle(\delta(x+d/2))^2.
\end{eqnarray} with $\hat{\rho}=|\Psi \rangle\langle \Psi|$
is the total density matrix. By analogy with equation (2) the
total distinguishability is then defined by
\begin{eqnarray}
K=\frac{||C_1|^2\langle\gamma_1|\gamma_1\rangle-|C_2|^2\langle\gamma_2|\gamma_2\rangle|}{|C_1|^2\langle\gamma_1|\gamma_1\rangle+|C_2|^2\langle\gamma_2|\gamma_2\rangle}.
\end{eqnarray}
Same as for equations (3-5) we can define the total probability to
detect a photon of (transverse) wave vector $k$ by
\begin{eqnarray}
P(k)=\textrm{Tr}[\hat{\rho}|k \rangle\langle k|]\propto 1 +
V\cos{(kx+\phi)},
\end{eqnarray} where the visibility $V$ is written
\begin{equation}
V=\frac{2|C_1|\cdot|C_2|\cdot|\langle\gamma_1|\gamma_2\rangle|}{|C_1|^2\langle\gamma_1|\gamma_1\rangle+|C_2|^2\langle\gamma_2|\gamma_2\rangle}.
\end{equation}
By combining  $V$ and $K$ we deduce immediately
\begin{eqnarray}
K^2+V^2=1-\frac{4|C_1|^2\cdot|C_2|^2\cdot(\langle\gamma_1|\gamma_1\rangle\langle\gamma_2|\gamma_2\rangle-|\langle\gamma_1|\gamma_2\rangle|^2)}{(|C_1|^2\langle\gamma_1|\gamma_1\rangle+|C_2|^2\langle\gamma_2|\gamma_2\rangle)^2}\leq
1,
\end{eqnarray}
where the last inequality results from the Cauchy-Schwartz
relation\\
$\langle\gamma_1|\gamma_1\rangle\langle\gamma_2|\gamma_2\rangle-|\langle\gamma_1|\gamma_2\rangle|^2\geq0$.\\
However, we can remark that  by tracing over the degrees of
freedom associated with the detector we didn't considered a which
path experiment but simply decoherence due to entanglement. In
order to actually realize such a which-path experiment we need to
calculate the joint probability associated with a recording of the
photon in the state $|x\rangle$ (or $|k\rangle$) in coincidence
with a measurement of the detector in the eigenstate
$|\lambda\rangle$ corresponding to one of its observable. These
joint probabilities read $P(x,\lambda)=\textrm{Tr}[\hat{\rho}|x
\rangle\langle x||\lambda\rangle\langle\lambda|]$ and
$P(k,\lambda)=\textrm{Tr}[\hat{\rho}|k\rangle\langle
k||\lambda\rangle\langle \lambda|$ with
\begin{eqnarray}
P(x,\lambda)\propto|C_1|^2|\langle\lambda|\gamma_1\rangle|^2(\delta(x-d/2))^2+|C_2|^2|\langle\lambda|\gamma_2\rangle|^2(\delta(x+d/2))^2\nonumber\\
P(k,\lambda)=\propto 1 + V_\lambda\cos{(kx+\phi_\lambda)}.
\end{eqnarray}
Indeed, the aim of such entanglement with a degree of freedom
$|\lambda\rangle$ (produced for example by inserting polarization
converters like quarter or half wave-plates just after the
apertures \cite{W}) is to generate a wave function
\begin{equation}
\psi_{\lambda}(x)\sim
C_{1,\lambda}\delta(x-d/2)+C_{2,\lambda}\delta(x+d/2)
\end{equation}
with either $C_{1,\lambda}$ or $C_{2,\lambda}$ (but not both)
equal to zero. A subsequent projection on $|\lambda\rangle$ will
reveal the path information. However, from the duality relation
given by equation (5) applied to $\psi_{\lambda}(x)$ it is now
obvious that we did not escape from the previous conclusion.
Indeed, while the photon was not destroyed by the entanglement
with the which-path detector, we unfortunately only obtained path
distinguishability ($K_{\lambda}=1$) at the
expense of losing the interference behavior ($V_{\lambda}=0$).\\
\indent From all these experiments, it is clear that the
discreteness of photon, and more generally of every quantum
object, is the key element to understand complementarity. This was
evident without entanglement, since the only way to observe a
particle is to destroy it. However, even the introduction of a
`which-path' quantum state $|\lambda\rangle$ does not change the
rule of the game, since at the end of journey we necessarily need
to project, that is to kill macroscopically, the quantum system.
This fundamental fact, was already pointed out many times by Bohr
in his writings when he considered the importance of separating
the macroscopic world of the observer from the microscopic quantum
system observed, and also when he insisted on the irreversible act
induced by the observer
on the quantum system  during any measurement process \cite{Bohr2}.\\
\indent  Let now return to the interpretation of Afshar \emph{et
al}. experiments. In the configuration with the lens and without
the grid, we have apparently a new aspect of the problem since the
fringes occur in a plane located before the imaging plane.
Contrarily to the which-path experiments above mentioned, where
the destructive measurements occurred in the interference plane,
we have a priori here the freedom to realize a
`fringes-interaction free-experiment' which aim is to observe the
fringes without detecting the particle in the back focal plane
whereas the destructive measurement will occur in the image plane
(i.e., in 1' or 2'). The role of the grid is expected to provide
such information necessary for the interference reconstruction.
Due to the absence of disturbance by the grid, Afshar \emph{et
al}. logically deduce that the field equals zero at the wires
locations. If we \emph{infer} the existence of an interference
pattern with visibility $V$ we must have
\begin{equation}
V=\frac{(I_{\texttt{max}}-I_{\texttt{min}})}{(I_{\texttt{max}}+I_{\texttt{min}})}=\frac{(I_{\texttt{max}}-0)}{(I_{\texttt{max}}+0)}=1,
\end{equation}
since $I_{\texttt{min}}=0$. This means that we can obtain the
value of the visibility only from the two assumptions  that (i)
the form of the profile should be a `cos' function given by
equation (4), and  that (ii) no photon have been absorbed by the
wires. Finally in this experiment, we record the photons in the
area 1' (or 2') and consequently we have at the same time the path
information. Importantly, following Afshar \emph{et al}. we here
only consider one image spot 1' or 2' (since each photon impinges
one only one of these two regions) and we deduce therefore $K=1$.
Together with the interference visibility $V=1$ this implies
\begin{equation}
K^2+ V^2=2,
\end{equation}
in complete contradiction with the duality-complementary bound
given by equation (6). In the previous analysis we only considered
the infinitely thin wires to simplify the discussion. Actually,
this is however the only experimental configuration in which the
Afshar experiment is easily analyzable since it is only in such
case that the duality relation can be defined. Indeed, scattering
by the wire always results into complicated diffraction pattern in
the image plane and the simple mathematical
derivation~\cite{Englert1,Englert2,Englert3,Englert4} leading to
equations 2, 5, and 6 is not possible. We will then continue to
consider the idealized case of the infinitely thin wires in the
rest of the paper since it is this ideal limit that the authors of
\cite{Afshar}wanted obviously to reach.
\section{The rebuttal: Inference and Complementarity}
\subsection{Duality again}
\indent There are several reasons why the analysis by Afshar
\emph{et al.} actually fails. First, from a mathematical point of
view it is not consistent to write $K^2+ V^2=2$. Indeed, in all
the experiments previously  discussed (excluding the Afshar
experiments) it was necessary to consider statistics on all the
recorded photons in order to observe either the interference or
the path information (in the case were entanglement was involved
only the photons tagged by $|\lambda\rangle$ have to be
considered). Same here, if one consider all the detected photons
one will deduce $K=0$ and equation (6) will be respected.
Actually, this results directly from the experimental method
considered by the authors of \cite{Afshar}. Indeed, if somebody is
accepting the existence of an interference pattern he or she needs
to know the complete distribution 1' \emph{and} 2' recorded in the
image plane. This is necessary in order to deduce that the wire
grid didn't caused any disturbances on the propagation. Indeed,
the disturbance could  have no consequence in 1' but yet have some
effects in 2'. Consequently, ignoring 2' does not allow us to
deduce that the experiment with the grid is interaction-free. For
this reason, it is unjustified to write $K=1$, that is to consider
only one half of the detected photon population, while we actually
need both pinhole images to deduce the value of $V$ (this is also
in agreement with the obvious fact that an interference pattern
requires the two apertures 1 and 2
opened for its existence).\\
\indent There is an other equivalent way to see why the choice
$K=0$ is the only one possible. Indeed, having measured in the
image plane the two distributions 1' and 2' with intensity
$|C_1|^2$ and $|C_2|^2$ we can, by applying the laws of optics,
propagate backward in time the two converging beams until the
interference plane (this was done by Afshar \emph{et al.}). In
this plane equation (4) and (5), which are a direct consequence of
these above mentioned optical laws, are of course valid . Since we
have $|C_1|^2=|C_2|^2$, we deduce (from equations (2) and (5))
that $K=0$ and $V=1$ in full agreement with the duality relation
(6). It is important to remark that since the phase of $C_1$ and
$C_2$ are not know from the destructive measurements in the image
plane, we cannot extrapolate the value of
$\chi=\arg{(C_1)}-\arg{(C_2)}$. However, the presence of the grid
give us access to this missing information since it provides the
points where $I(k)=0$ (for example if $I(\pi/d)=0$ then
$\chi=2\pi\cdot N$ with $N=$0, 1, ...). We can thus define
completely the variable $V$ and $\chi$ without having measured any
photon in the Fourier plane. It is also clear, that this would
have been impossible if the duality condition $K^2+ V^2=1$ was not
true since this relation is actually a direct consequence of the
law of optics used in our derivations as well as in the one by
Afshar \emph{et al.}.\\
\indent To summarize the present discussion, we showed that Afshar
\emph{et al.} reasoning is obscured by a misleading interpretation
of the duality relation given by equation (6).  We however think
that this problem is not so fundamental for the discussion of the
experiment. Actually, we can restate the complete reasoning
without making any reference to this illusory  violation of
equation (6). After doing this we think  that the error in the
deductions by Afshar \emph{et al.} should become very clear. Let
then restate
the story: \\
\indent A) First, we record individuals photons in the regions 1'
and 2'. We can then keep a track or a list of each detection
event, so that, for each photon, we can define its `path'
information. However, this individual property of each photon is
not entering in conflict with the statistical behavior, which in
the limit of large number, give us the two narrow distribution in
1' and 2'. That is, the value $K=0$ is not in conflict with the
existence of a which path information associated with each photon.
This situation differs strongly from the previously which path
experiments where the path detection, or tagging, is done
\emph{before} the interference plane. As we explained  before in
these experiments the value $K=1$ was a necessary consequence of
the preselection procedure done on the photon population.  This
point also means that we have to be very prudent when we use the
duality relation in experimental situations different from the
ones for which a
consensus has already been obtained.\\
\indent B) Second, we apply the laws of optics backward in time to
deduce the value of the visibility $V$. Inferring the validity of
such optical laws we can even reconstruct completely the
interference profile thanks to the presence of wire grid.\\
\indent C) Finally, we can check that indeed $K^2+ V^2=1$ in
agreement with the duality relation.\\
\indent Having elucidated the role of the duality relation, the
question that we have still to answer is what are the implications
of this experiment for complementarity. What has indeed been shown
by Afshar \emph{et al.} is that each photon detected in the image
plane is associated with a wave behavior since none of them
crossed the wires. Using the laws of optics backward in time allow
us to deduce the precise shape of intensity profile in the back
focal plane but this is a theoretical inference and actually not a
measurement. We will now show that this is the key issue.
\subsection{Classical versus quantum inferences}
\indent In classical physics, such an inference (i.e., concerning
interference) is of no consequence since we can always, at least
in principle, imagine a test particle or detector to check the
validity of our assumptions concerning the system. However, in
quantum mechanics we are dealing with highly fragile systems and
this modify the rules of the game.
\begin{figure}[h]
\begin{center}
\includegraphics[width=10cm]{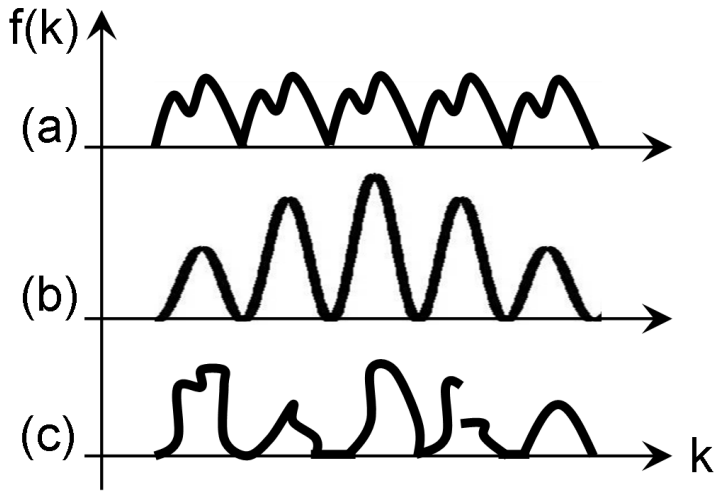}
\end{center}
\caption{Different possible intensity profiles in the Fourier
plane. Each profile $f(k)$ obeys to the condition $f(k)=0$ on the
wires. (a) A continuous periodic function. (b) The diffractive
interference profile predicted by quantum mechanics. (c) A
discontinuous profile intensity. Each profile is `apriori'
equiprobable for an observer which has no knowledge in optics and
quantum mechanics. }
\end{figure}
In quantum  mechanics it is common to say that the wave function
represents the catalog of all the potentiality accessible to the
system. Due to the very nature of this theory there are however
some (complementary) pages which can not be read at the same time
without contradictions. In the Afshar experiment, we do not have
indeed the slightest experimental proof that the observed photons
did participate to the `cos' interference pattern given by
equations (3) and (4). Furthermore, by detecting the photons in
the image plane, we only know from the experiment that the photons
never crossed the wires but this is not sufficient to rebuild
objectively the
complete interference pattern.\\
\indent We can go further in this  direction by  using information
theory. Indeed, from the point of view of the information theory
of Gibbs~\cite{Gibbs}, Shannon~\cite{Shannon}, and
Jaynes~\cite{Jaynes}, every interference patterns, such that
$I(k)=0$ on the wires, are equiprobable (see Fig.~2). However,
there are an infinity of such profiles, so that our information is
rather poor. More precisely, let write $\rho[f(k)]$ the functional
giving the density of probability associated with the apriori
likelihood of having the interference profile $f(k)$ located in an
infinitely small (functional) volume $\mathcal{D}[f(k)]$. We write
$\Sigma[f(x)]$ the space of all this interference profiles obeying
to the condition $f(k)=0$ on the wires. We have thus
$\rho[f(k)]=1/\Sigma$ (equiprobability) for the function $f$
contained in $\Sigma$, and $\rho[f(k)]=0$ for the function outside
$\Sigma$ (that are functions which do not satisfy the requirements
$f(k)=0$ on the wires). The Shannon
entropy\cite{Gibbs,Shannon,Jaynes} $S[f(x)]$ associated with this
distribution is given by
\begin{eqnarray}
S[f(x)]=-\int_{(\Sigma)}\mathcal{D}[f(k)]\rho[f(k)]\ln{(\rho[f(k)])}
\nonumber\\
=\ln{(\Sigma[f(k)])}\rightarrow +\infty,
\end{eqnarray}
which expresses our absence of objective knowledge concerning
$f(k)$. In this reasoning, we used the concept of probability
taken in the Bayesian sense, that is in the sense of
decision-maker theory used for example by poker players. For an
observer which do not have any idea concerning quantum mechanics
and the laws of optics, this `subjective' equiprobable guess is
the most reasonable if he wants only to consider the photons he
actually detected. Of course, by considering a different
experiment, in which the photons are recorded in the Fourier
plane, the observer might realize what is actually the
interference pattern.  However (and this is essential  for
understanding the apparent paradox discussed in reference~1) it
will be only possible by considering different recorded photons in
full agreement with
the philosophy of the principle of complementarity.\\
\indent Let now summarize a bit our analysis. We deduced that in
the experiment discussed in \cite{Afshar} the photons used to
\emph{measure} objectively the interference pattern and so to
calculate the visibility $V=1$ are not the same than those used to
\emph{measure} the distribution in the image plane and calculate
the distinguishability $K=0$. This is strictly the same situation
than in the original two-holes experiment already mentioned. It is
in that sense that the relationship (6) represents indeed a
particular formulation of complementarity
\cite{Englert1,Englert2,Englert3,Englert4}. Actually (as we
already commented before) the value $V=1$ obtained in
\cite{Afshar} does not result from a measurement but from an
extrapolation. Indeed, from their negative measurement Afshar et
al. recorded objectively $I_{min}=0$. If we suppose that there is
a hidden sinusoidal interference pattern in the plane of the wires
we can indeed write
\begin{equation}
V=\left(I_{max}-I_{min}\right)/\left(I_{max}+I_{min}\right)=I_{max}/I_{max}=1.
\end{equation}
However to prove experimentally that such sinusoidal interference
pattern actually exists we must definitively record photons in the
rest of the wires plane. This is why the experiment described in
\cite{Afshar} does not constitutes a violation of
complementarity.\\
\indent It is finally interesting to remark that similar analysis
can be easily done already in the Young two-holes experiment.
Indeed, suppose that we record the photon interference fringes
after the holes. We can thus measure $V=1$. However, if we suppose
that the sinusoidal oscillation of the intensity results from the
linear superposition of waves coming from holes 1 and 2 then from
equation 5 we deduce $|C_1|^{2}+|C_2|^{2}-2|C_1||C_2|=0$ i.~e.,
$|C_1|=|C_2|$. From equation 2 this implies $K=0$. Reasoning like
Afshar \emph{et al.} we could be tempted to see once again a
violation of complementarity since we deduced the
distinguishability without disturbing the fringes! However, we
think that our previous analysis sufficiently clarified the
problem so that paradoxes of that kind are now naturally solved
without supplementary comments.
\subsection{The objectivity of trajectory in quantum mechanics}
At the end of section 2.1 we shortly pointed that the concept of
trajectory is a key issue in the analysis of the experiment
reported in reference~1. This was also at the core of most
commentaries (e.g., references \cite{B,C,D,E,F,G,H,I,J})
concerning the work by Afshar \emph{et al.}. As a corollary to the
previous analysis we will now make a brief comment concerning the
concept of path and trajectory in quantum mechanics since we think
that a lot of confusion surrounds this problem. This is also
important because Afshar \emph{et al.} claimed not only that they
can circumvent complementarity but that additionally they
determine the \emph{path} chosen by the particle. Following here
an intuitive assumption they accepted that with the two pinholes
open a photon trajectory (if trajectory there is) connects
necessarily a pinhole to its optical image like it is in
geometrical optics. They called that intuition (probably in
analogy with what occurs in classical physics) a `consequence of
momentum conservation'. However, the meaning of momentum and
trajectory is not the same in quantum and classical mechanics.
Actually as it was realized by several physicists the connection 1
to 1' and  2 to 2' is a strong hypothesis which depends of our
model of (hidden) reality and which can not in general be
experimentally tested (read for example " Surrealistic Bohm
trajectories" \cite{Scully2} and also \cite{Hiley})].
\begin{figure}[h]
\begin{center}\includegraphics[width=10cm]{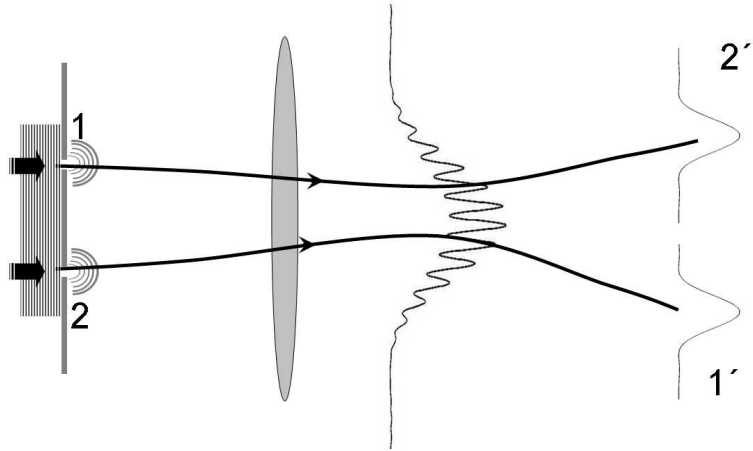}\end{center}
\caption{Illustration of the counterintuitive paths followed by photons if we accept the ontological interpretation given by de Broglie and Bohm.
The photons coming from aperture 1 or 2 reach the `wrong' detector 2' or 1'. }
\end{figure}Actually
nothing in this experiment with two holes forbids a photon coming
from one pinhole to go in the \emph{wrong} detector associated
with the second pinhole. This is the case for example in the
\emph{hidden variable} theory of de Broglie-Bohm in which every
photons coming from the aperture 1 (respectively 2) is reaching
the wrong image spot 2'(respectively 1') \cite{Wesley,Dewdney} as
shown in figure 3. This is counter intuitive but not in
contradiction with experiments since we can not objectively test
such hidden variable model \cite{Scully2,Hiley}. In particular
closing one pinhole will define unambiguously the path followed by
the particle. However this is a different experiment and the model
shows that the trajectories are modified (in general non locally)
by the experimental context. The very existence of a model like
the one of de Broglie and Bohm demonstrates clearly that in the
(hidden) quantum world  a trajectory could depend of the complete
context of the experiment. For this reason we must be very prudent
and conservative when we interpret an experiment: Looking the
image of a pinhole recorded in a statistical way by a myriad of
photon will not tell us from which pinhole an individual photon
come from but only how many photons crossed this pinhole. In
counterpart of course we can not see the fringes and the
complementarity principle of Bohr will be, as in every quantum
experiment, naturally respected. It is thus in general dangerous
to speak unambiguously of a which path experiment and this should
preferably be avoided from every discussions limited to empirical
facts. As claimed by Bohr the best empirical choice is in such
conditions to accept that \emph{it is wrong to think that the task
of physics is to find out how Nature is. Physics concerns what we
can say about Nature} \cite{Bohr2}.
\section{CONCLUSION} To
conclude, in spite of some claims we still need  at least two
complementary experiments in order to exploit the totality of the
phenomenon in Young-like interferometers. Actually, as pointed out
originally by Bohr, we can not use information associated with a
same photon event to reconstruct in a statistical way (i.e. by a
accumulation of such events) the two complementary distributions
of photons in the image plane of the lens and in the interference
plane. The presence of the wires inserted in reference~1 does not
change anything to this fact since the information obtained by
adding the wires is too weak and not sufficient to rebuild
objectively (i.~e.~, unambiguously from experimental data) the
whole interference pattern. The reasoning of Afshar \emph{et al.}
is therefore circular and the experiment is finally in complete
agreement with the principle of complementarity.
\section{Acknowledgments}
I thank  M.~Gotlieb and G. Perec~\cite{Perec} for motivating (implicitly and subconsciously) the mere existence of this work.


\begin{thebibliography}{99}
\bibitem{Remark1}
In the actual experimental setup considered in \cite{Afshar} the
lens is located far away from the two pinholes so that it is in
practice equivalent to observe the fringes in front of the lens or
in its back focal plane.
\bibitem{Remark2}
The wording `path' is here chosen in agreement with the current
usage in quantum mechanics text-books. However, its meaning should
not be taken too cautiously since there are still vivid debates
concerning what should be call a path or a trajectory in quantum
mechanics. This is particularly true if one compare the position
of the Copenhagen school on this subject~\cite{Heisenberg3,Bohr1}
to the answer given by Bohmian
mechanics~\cite{Scully2,Vaidman,BohmB}. See also the section 3.3.
\end{thebibliography}
\end{document}